\documentclass[pra,showpacs,twocolumn,floats,epsf]{revtex4}
\usepackage{graphicx}
\usepackage{dcolumn}
\usepackage{epsf}
\usepackage{amsmath}

\DeclareSymbolFont{AMSb}{U}{msb}{m}{n}
\DeclareMathSymbol{\N}{\mathbin}{AMSb}{"4E}
\DeclareMathSymbol{\Z}{\mathbin}{AMSb}{"5A}
\DeclareMathSymbol{\R}{\mathbin}{AMSb}{"52}
\DeclareMathSymbol{\Q}{\mathbin}{AMSb}{"51}
\DeclareMathSymbol{\I}{\mathbin}{AMSb}{"49}
\DeclareMathSymbol{\C}{\mathbin}{AMSb}{"43}

\def\openone{\leavevmode\hbox{\small1\kern-3.3pt\normalsize1}}

\newcommand{\be}{\begin{equation}}
\newcommand{\ee}{\end{equation}}
\newcommand{\bea}{\begin{eqnarray}}
\newcommand{\eea}{\end{eqnarray}}

\newcommand{\var}{{\rm var}\,}
\newcommand{\tr}{{\rm tr}\,}

\begin{document}

\title{Controlled decoherence in a quantum L\'evy kicked
rotator}

\author{Henning Schomerus}
\affiliation{Department of Physics, Lancaster University,
Lancaster, LA1 4YB, UK}
\author{Eric Lutz}
\affiliation{Department of Physics, University of Augsburg,
D-86135 Augsburg, Germany}

\date{\today}

\begin{abstract}
We develop a theory describing the dynamics of quantum kicked
rotators (modelling cold atoms in a pulsed optical field) which
are subjected to combined amplitude and timing noise generated by
a renewal process (acting as an engineered reservoir). For
waiting-time distributions of variable exponent (L\'evy noise), we
demonstrate the existence of a regime of nonexponential loss of
phase coherence. In this regime, the momentum dynamics is
subdiffusive, which also manifests itself in a non-Gaussian
limiting distribution and a fractional power-law decay of the
inverse participation ratio. The purity initially decays with a
stretched exponential which is followed by two regimes of
power-law decay with different exponents. The averaged logarithm
of the fidelity probes the sprinkling distribution of the renewal
process. These analytical results are confirmed by numerical
computations on quantum kicked rotators subjected to noise events
generated by a Yule-Simon distribution.
\end{abstract}


\maketitle

\section{Introduction}

The coupling of an open quantum system to an external reservoir
induces a dynamical loss of phase coherence that results in an
irreversible destruction of interference effects between states of
the system \cite{zur03,giu96}. This environment-induced
decoherence efficiently suppresses  macroscopic superpositions of
states and therefore plays a central role in our understanding of
the transition from quantum to classical physics. An important and
often annoying feature of decoherence for quantum-mechanical
applications is that it typically occurs very rapidly with an
exponential decoherence factor, ${\cal D}(t) = \exp(- t/t_c)$,
where $t_c$ is the short coherence time \cite{zur03, giu96}. In
some cases, however, the coherence time can be controlled by
properly engineering the reservoir and/or the interaction between
the system and the reservoir \cite{poy96}. The feasibility of such
schemes has been experimentally demonstrated in a variety of
systems that include cavity QED \cite{bru96}, ion traps
\cite{mya00} and cold atom experiments \cite{amm98,kla98,osk03}.
Besides their obvious technological implications, these
experiments provide essential tests of the general concept of
environment-induced decoherence.

The atom-optical experiments \cite{amm98,kla98,osk03}
 are designed to realize the quantum-kicked rotator
\cite{chi78,graham,moo95}, a paradigm of complex chaotic dynamics
which can be simply seen as a particle moving on a ring and
periodically kicked in time. For sufficiently strong kicking
strength, the classical motion of a kicked rotator is chaotic and
the evolution in momentum space is diffusive. In the quantum
regime, however, momentum diffusion is asymptotically suppressed
due to dynamical localization \cite{fis82,izr90}.  In the recent
atom-optical reservoir-engineering experiments, the loss of
coherence has been induced by amplitude noise \cite{kla98} and
timing noise \cite{osk03}. In the first case, the amplitude of the
periodic kicks is randomly modulated in a controlled manner,
whereas in the second case the period between successive kicks is
randomly varied. In both situations, dynamical localization is
extenuated leading to a non-vanishing quantum momentum diffusion
with a renormalized diffusion coefficient, in agreement with
theoretical predictions \cite{ott84,coh91,kla98}. These
experiments enjoy a high degree of control and tunability, which
makes the atom-optical setup ideally suited to investigate the
decoherence-induced quantum-to-classical crossover of complex
quantum systems \cite{Tworzydlo}.

One point is especially noteworthy: reservoir engineering has so
far been used to tune the coherence time, but not the exponential
time dependence of the decoherence factor ${\cal D}(t)$ itself.
Our aim in this paper is to describe a reservoir coupling scheme
for the cold atom experiments which allows to modify the time
dependence of  ${\cal D}(t)$ and slow down the loss of phase
coherence in a controlled way. To this end, we superpose the
regular periodic pulses with noise events occurring with randomly
modulated time intervals $\tau$ ({\em i.\,e.}, we combine
amplitude with timing noise). We generate these random time
intervals by a renewal process with a waiting-time distribution
$w(\tau)$. Of special interest is the case where this distribution
asymptotically behaves as a power law, $w(\tau)\propto
\tau^{-1-\alpha}$. By tuning the exponent $\alpha$ of this
so-called L\'evy noise \cite{shl93}, we are able to change the
mean waiting time between successive noise events, given by
$\overline\tau=\int_0^\infty d\tau \,\tau \,w(\tau)$, and
therefore can smoothly interpolate between a fully coupled
situation where most kicks are perturbed by the noise (large
$\alpha$) to an almost isolated situation where most kicks are not
perturbed by the noise (small $\alpha$).

Divergent moments are a hallmark of L\'evy statistics and the mean
waiting time $\overline\tau$ becomes infinite when $\alpha \leq
1$. In this case the noise is nonstationary and nonergodic
\cite{lut04,mar05}. We show that this type of noise  modifies the
decoherence of the atoms in a striking manner: The exponential
time dependence of the decoherence factor ${\cal D}(t)$ is
replaced by a Mittag-Leffler function \cite{MLfunction}, which
starts out as a stretched exponential and asymptotically decays as
a power-law. For this functional form, the coherence time is
ill-defined.

In Ref.\ \cite{sch07} we briefly discussed one manifestation of
the modified decoherence pattern, namely a subdiffusive momentum
spreading of the atoms. In this work, we present a theory that
links the subdiffusive behavior to the nonstationarity of the noise.
This requires us to extend the previous theories of Refs.\
\cite{ott84,coh91} beyond the perturbative short-time regime
and to account for the additional timing noise from the renewal
process. We also discuss the limiting distribution function and
the inverse participation ratio, which provide additional
information about the dynamics in momentum space. As
alternative measures of coherence, we then consider the purity
and the fidelity of atoms subjected to different realizations of
the noise, which can be probed in an echo experiment
\cite{jalabert,echo}. For nonexponential decoherence, we
find that the purity initially decays with a stretched
exponential, which is followed by two regimes of power-law
decay with different exponents. The decay of the averaged
logarithm of the fidelity is related to the sprinkling
distribution of the renewal process. Throughout the paper our
findings are illustrated by comparison to the results from a
numerical implementation of the kicked rotator.

The structure of this paper is as follows. Section \ref{sec:model}
describes the quantum kicked rotator and the environmental
coupling scheme of combined amplitude and timing noise generated
by a renewal process. In Sec.\ \ref{sec:theory} we derive the
nonperturbative expression for the decoherence factor and evaluate
it for L\'evy noise. In the case of nonstationary noise with
$\alpha \leq 1$, the decoherence function is nonexponential.
Section \ref{sec:momentum} discusses how the modified decoherence
affects the momentum spreading, while Sec.\ \ref{sec:purity}
contains the discussion of the purity and fidelity. The results of
this paper are summarized in Sec. \ref{sec:conclusions}.

\section{The L\'evy kicked rotator\label{sec:model}}
\subsection{Model}

The motion of atoms exposed to pulsed
standing-wave potentials can be mapped onto a quantum kicked
rotator \cite{chi78,graham,moo95}, i.e., a point particle
 moving  on a ring and subjected to periodical kicks. In
suitable units, the Hamiltonian of the kicked rotator takes the
form
\begin{equation}
\label{eq1} H=\frac{p^2}{2}+\sum_{n=-\infty}^\infty
K_n\cos\theta\,\delta(t-n+0^+),
\end{equation}
where the kicking potential depends on the $2\pi$-periodic
rotation angle $\theta$ and the kicking amplitude $K_n$ can in
general be kick dependent.

From kick to kick, the stroboscopic classical dynamics of the
kicked rotator is generated by the map
\begin{eqnarray} \label{eq:pmap} &&p(t+1)=p(t)+K_t\sin\theta,\\
&&\theta(t+1)=\theta(t)+p(t+1).
\end{eqnarray}
Starting from a given initial state $\psi(0)$, the corresponding
quantum dynamics $\psi(t+1)=F(K_t)\psi(t)$ is generated by the
Floquet operator
\begin{equation}
\label{eq:floquet}
 F(K_t)=\exp(-i\hbar^{-1}\hat
p^2/2)\exp(-i\hbar^{-1}K_{t}\cos\hat \theta).
\end{equation}
In both cases, the dynamics consists of a kick in which the
momentum changes by $K_t\sin\theta$, followed by a free rotation
in which the rotation angle $\theta$ increases by $p$.

In the absence of noise, the dynamics of the kicked rotator is
entirely controlled by the constant  parameter $K_n=K$. For
$K\gtrsim 5$, the  classical dynamics is chaotic  and the growth
of the  momentum is  on average diffusive with a variance  $\var p
(t) \simeq D_{\rm cl} t$. The classical diffusion constant is
given by $D_{\rm cl}\simeq K^2/2$ \cite{rechester}. By contrast,
in the quantum regime, the kicked rotator exhibits dynamical
localization, an interference phenomenon akin to Anderson
localization in disordered solids, which manifests itself in an
exponentially decaying envelope of the quasienergy eigenstates
$F(k)$ of the Floquet operator \cite{fis82}. As a result, momentum
diffusion is suppressed and the variance
\begin{equation}
\var p_0(t) \simeq D^* t^*[1-\exp(-t/t^*)] \label{eq:varp0}
\end{equation}
saturates after the quantum break time $t^* \simeq D^*/ \hbar^2$
\cite{izr90}. The constant  $D^*$ is of order of the classical
diffusion constant $D_{\rm cl}$, but is subject to quantum
corrections \cite{shep}.

The transition between the quantum and the classical behavior
of the rotator can be induced by subjecting the kicked particle
to additional random kicks. In the case of amplitude noise
\cite{kla98}, the kicking amplitude is explicitly $n$-dependent,
$K_n=K+k_n$, where the perturbations $k_n$ are random numbers.
In the presence of conventional timing noise \cite{osk03} the
period between the kicks (here set to unity) is slightly
modulated.

\begin{figure}
\epsfxsize=\columnwidth \epsffile{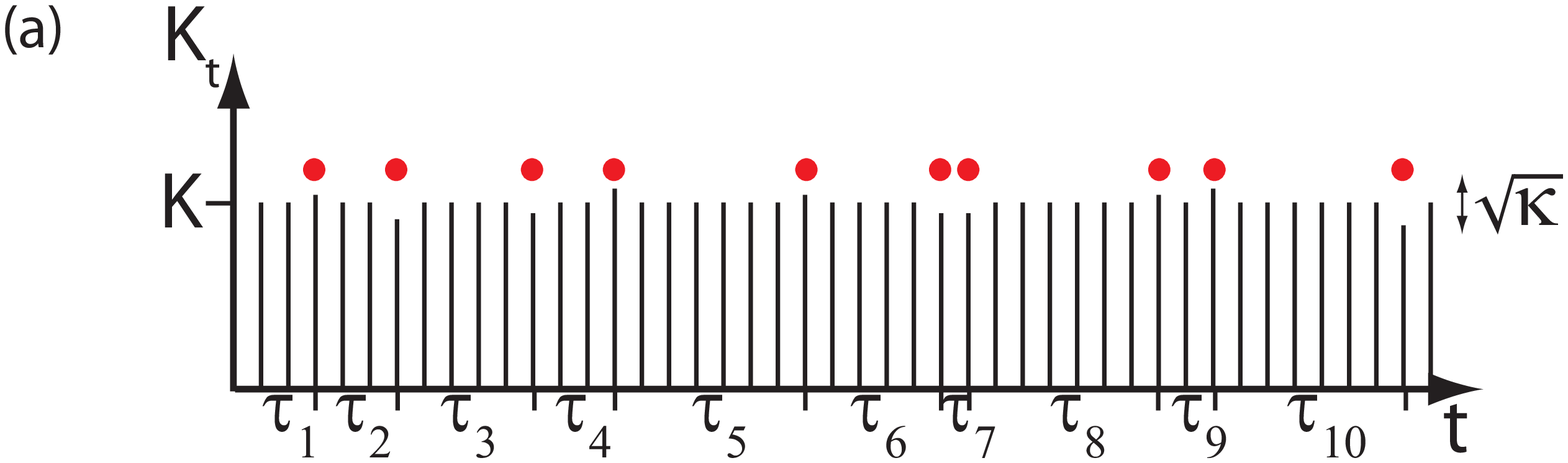}
\epsfxsize=\columnwidth \epsffile{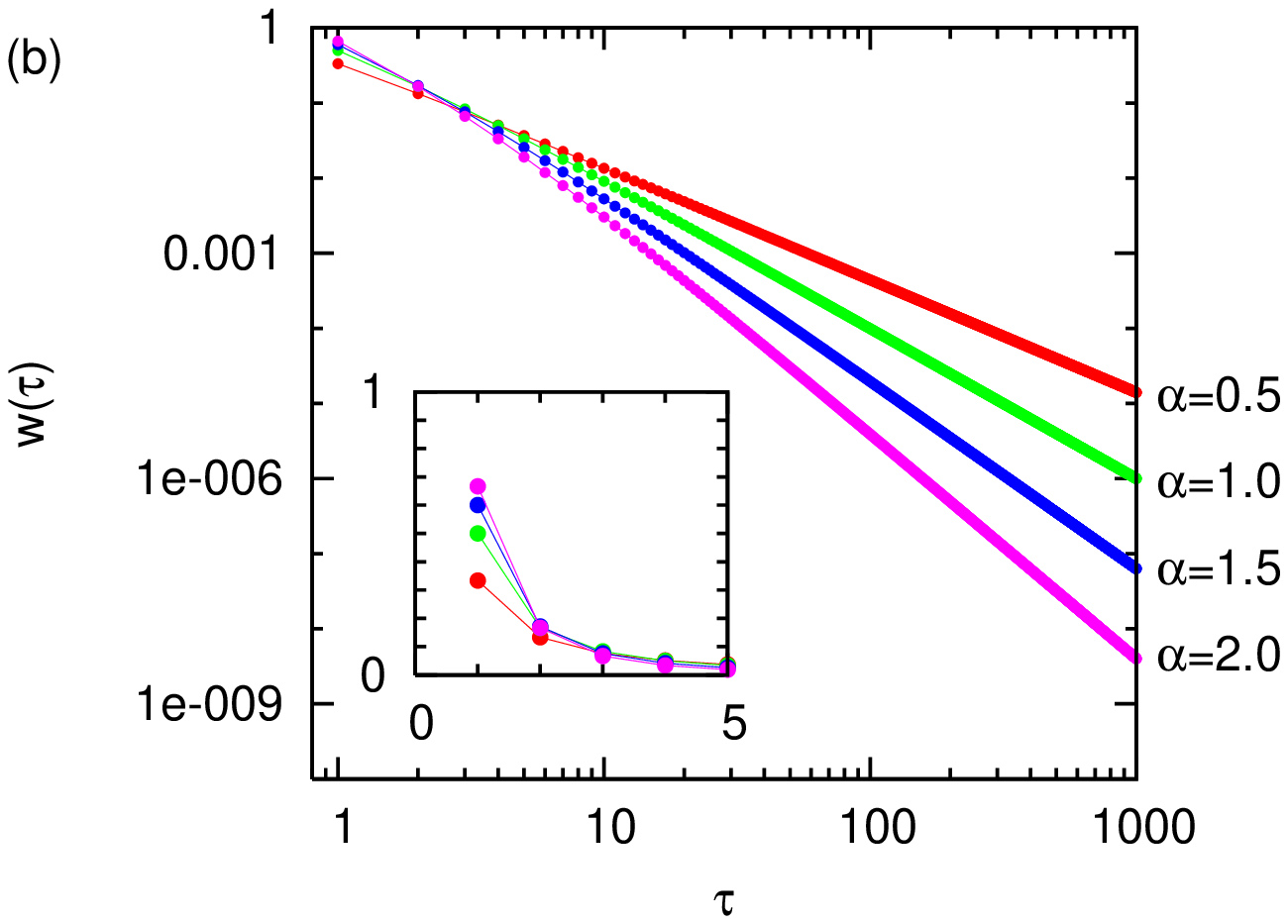}
\caption{\label{fig:1}(color online). (a) Schematic illustration
of the kicking sequence in the L\'evy kicked rotator. The period
of the kicks is set to unity. The timing of the noise is generated
in a renewal process, which selects the integer-valued waiting
times $\tau_n$ from a distribution $\omega(\tau)$. (b)
Waiting-time distributions of the Yule-Simon form (\ref{eq:yule})
for different values of the parameter $\alpha$. The distributions
display power-law tails, specified in Eq.\ (\ref{eq:yuleasym}),
and hence generate L{\'e}vy noise of variable exponent. The lines
are a guide to the eye.}
\end{figure}

In this paper we consider an unconventional combination of
amplitude and timing noise: kicks always appear at periodic
instances and the amplitude of some of the kicks are perturbed.
However, the (integer-valued) time intervals between the noise
events are generated by a renewal process with waiting-time
distribution $\omega(\tau)$ (see Fig.\ \ref{fig:1} for a schematic
illustration). The timing of the ${\cal N}$'th noise event (${\cal
N}=1,2,3,\ldots)$ is hence given by
\begin{equation}\label{eq:randomtime}
t_{\cal N}=\sum_{n=1}^{\cal N}\tau_n,
\end{equation}
where each waiting time $\tau_n$ is drawn independently from
the same probability distribution $\omega(\tau)$. The strength
of the individual noise events will be characterized by the
variance $\overline{k_{t_{\cal N}}^2}=\kappa$ of the perturbed
kicks, whereas on average $\overline{k_{t_{\cal N}}}=0$.

The usual white amplitude noise is obtained for a waiting-time
distribution $\omega(\tau)=\delta_{\tau,1}$, so that every
kick is perturbed. A more intriguing case is a  waiting-time
distribution which asymptotically behaves as a power
law (L\'evy noise),
\begin{equation}\label{eq:w}
w(\tau)\sim c\,\tau^{-1-\alpha}, \end{equation} where $c$ is a
constant. For $\alpha<1$ the mean waiting time $\overline\tau$
diverges. Unlike the familiar  white amplitude noise, L\'evy noise  \cite{genericlevyreferences} generates
a nonstationary process with unconventional statistical
features that are reviewed in the following subsection.

\subsection{Statistical properties of L\'evy noise\label{sec:levy}}

The  properties of noise generated by a renewal process can be
conveniently quantified by analyzing the number of noise events
${\cal N}(t',t'')$ within an interval $[t'',t')$. The number
${\cal N}(t,0)$ is obtained by inverting the random time
$t_{\cal N}$, and hence is known as the {\em inverse random
time} of the process \cite{MLpaper}. On average, the rate of change $f(t)=\partial_t\overline{{\cal
N}(t,0)}$ of the inverse random time gives the probability that
there is a noise event at time $t$. This rate is called the
sprinkling distribution and is directly related to the
waiting-time distribution $\omega(\tau)$ via a Laplace transformation,
\begin{equation}\label{eq:fw}
\tilde f(u)=\int_0^\infty dt\, e^{-ut}f(t) =\frac{\tilde
w(u)}{1-\tilde w(u)}.
\end{equation}
We  denote throughout the article Laplace-transformed
functions by a tilde.

For small values of $u$, the power-law waiting-time distribution
(\ref{eq:w}) delivers
\begin{eqnarray}
&&\tilde w(u)\sim 1-u\overline{\tau}
\qquad\qquad\qquad\qquad(\alpha>1),
\\
&&\tilde w(u) \sim
1-u^\alpha\frac{c\pi}{\Gamma(\alpha+1)\sin\pi\alpha}
\quad(\alpha<1). \label{eq:wu}
\end{eqnarray}
For $\alpha>1$,   it then follows from Eq.\ (\ref{eq:fw}) that  the
sprinkling distribution takes a constant value at large times,
\begin{equation}
f(t)\sim \frac{1}{\overline{\tau}} \quad(\alpha>1).
\end{equation}
As a result, the noise asymptotically becomes
stationary in this case. However, for power-law waiting-time distributions with
$\alpha<1$,  the sprinkling distribution asymptotically
decays over time according to
\begin{equation}\label{eq:falphaleq1}
f(t)\sim t^{\alpha-1}\alpha\sin(\pi\alpha)/(\pi c)
\quad(\alpha<1).
\end{equation}
Due to the explicit time dependence of $f(t)$, the noise here remains nonstationary. We note that nonstationary also  implies nonergodicity, since time averages cannot be replaced by ensemble averages
\cite{mar05}. A direct consequence of Eq.~(\ref{eq:falphaleq1}) is that the mean inverse random time increases sublinearly,
\begin{equation}\label{eq:nalphaleq1}
\overline{{\cal N}(t,0)}\sim t^{\alpha}\sin(\pi\alpha)/(\pi c)
\quad(\alpha<1).
\end{equation}

More detailed information about  the statistical properties of the
renewal process can be obtained by analyzing the complete
distribution function ${\cal P}({\cal N};t',t'')$ of the number of
noise events. The latter is most conveniently characterized by the
moment-generating function \begin{equation} \label{eq:mdef} {\cal
M}(z;t',t'')=\overline{\exp[z{\cal
N}(t',t'')]}=\sum_{n=0}^{\infty} e^{z{\cal N}} {\cal P}({\cal
N};t',t'').\end{equation} For general time arguments $t'$ and
$t''$, this function can be expressed as
\begin{eqnarray}
{\cal M}(z;t',t'')&=&{\cal M}(z;t',0)\nonumber
\\
&&-(e^z-1)\int_0^{t''}ds\,f(s){\cal M}(z;t'-s,0),\nonumber
\\
\end{eqnarray}
so that it suffices to study the moment-generating function of the
inverse random time ${\cal N}(t,0)$.

We first relate the distribution function ${\cal P}({\cal N};t,0)$
of the inverse random time to the probability distribution
$P(t_{\cal N};{\cal N})$ of the ${\cal N}$th random time, which
succeeds via the intuitive expression
\begin{equation}\label{eq:pp}
{\cal P}({\cal N};t,0)=\int_{0}^{t}dt'\,[P(t';{\cal N})-P(t';{\cal
N}+1)].
\end{equation}
Since Eq.\ (\ref{eq:randomtime}) involves a sum of ${\cal N}$
independent waiting times, the Laplace transform of $P(t_{\cal N};{\cal N})$  factorizes,
\begin{equation}
\tilde P(u;{\cal N})=[\tilde w(u)]^{\cal N} ,
\end{equation}
 and Eq.\ (\ref{eq:pp}) thus yields
\begin{equation} \widetilde {\cal
P}({\cal N};u,0)=\int_0^\infty dt\, e^{-u t} {\cal
 P}(z;t,0)
=\frac{1}{u}[1-\tilde w(u)][\tilde w(u)]^{\cal N}.
\end{equation}
(As indicated, the Laplace transformation is taken with respect to
the later time argument). By  using the definition (\ref{eq:mdef})
of the moment-generating function ${\cal M}(z;t,0)$, we find that
its  Laplace-transform can be expressed  as a function of the
sprinkling distribution,
\begin{equation} \widetilde {\cal
 M}(z;u,0)= \frac{1}{u}\frac{1}{1-\tilde f(u)(e^z-1)}.
\end{equation}
One should note the explicit periodicity,  $\widetilde {\cal
 M}(z;u,0)=\widetilde {\cal
 M}(z+2\pi i;u,0)$, which is a consequence of the discreteness of
 the inverse random time.

When next going back to the time domain, we again have to
distinguish between asymptotically stationary and nonstationary noise.
In the former case $(\alpha>1)$, the moment-generating function is
asymptotically exponential,
\begin{equation} \label{eq:statdecay} {\cal
M}(z;t,0)\sim\exp[(e^z-1)t/\overline{\tau}] \quad(\alpha>1).
\end{equation}
Equation (\ref{eq:statdecay}) characterizes the counting statistics of a Poisson
process with mean rate $\overline{\tau}^{-1}$. In the
nonstationary case $\alpha<1$, inserting the asymptotic
behavior (\ref{eq:wu}) produces a moment-generating function of
the form
\begin{equation} \label{eq:m1} {\cal
M}(z;t,0)\sim E_\alpha[t^\alpha(e^z-1)
\Gamma(1+\alpha)(\sin\pi\alpha)/(\pi c)] \quad(\alpha<1),
\end{equation}
where $E_\alpha$ is the Mittag-Leffler function
\cite{MLfunction}, defined by
\begin{equation} E_\alpha(z)=\sum_{n=0}^\infty
\frac{z^n}{\Gamma(\alpha n+1)}.
\end{equation}
Expression (\ref{eq:m1}) becomes exact for all times when one
passes to continuous inverse times ($e^z-1 \to z$) and
considers the particular case of a scale-free L{\'e}vy process
\cite{MLpaper}. The moment-generating function then starts out
as a stretched exponential, ${\cal M}(z;t,0) \simeq
\exp\{t^\alpha /[\Gamma(-\alpha) c t_c]\}$, while for large
times it crosses over to the power law  ${\cal M}(z;t,0)\simeq
(c t_c/\alpha) t^{-\alpha}$.  For more general waiting-time
distributions, a reasonable approximation is to adapt the time
argument by using the mean inverse random time, such that
\begin{equation} \label{eq:ma} {\cal
M}(z;t,0)\approx E_\alpha\left[(e^z-1)
\Gamma(1+\alpha)\overline{{\cal N}(t,0)} \right] \quad(\alpha<1).
\end{equation}
This results in the same power-law asymptotics for large times,
and accounts for transient behavior in the initial decay.

\subsection{Numerical implementation}

Throughout this paper we compare  analytical results to results
obtained from a numerical implementation of the L\'evy kicked
rotator. In the present section, we describe the parameters used for these
computations.

Dynamical localization is expected to be the strongest when
$\hbar/(2\pi)$ is approximating a quadratic irrational
$\omega=\frac{1}{2}(\sqrt{a^2+4}-a)$. In order to render the
Hilbert space finite we choose $a=24$, set $\hbar=2\pi M/N$,  and
use the third convergent of the continued-fraction representation
of the quadratic irrational,
$\frac{1}{24+}\frac{1}{24+}\frac{1}{24}$. This gives $M=577$,
$N=13872$.

The integer $N$ is the Hilbert space dimension, which now is
large but finite, corresponding to a quantized momentum
$p=\hbar l$, $l=0,1,2,\ldots N-1$.  Similarly, in position
space the momentum quantization carries over to discretized
positions $\theta=2\pi l/N$. The integer $M$ determines the
quantum-mechanical periodicity in momentum space ($p$ and
$p+2\pi M$ are equivalent). Each classical period of $2\pi$
covers $N/M\approx 24$ discretized momentum values.

In all our computations, the particle is initially prepared in
the zero-momentum state $\psi(0)=|p=0\rangle$. The  subsequent
propagation is obtained by successive application of the
Floquet operator (\ref{eq:floquet}), which can be broken down
into the application of diagonal matrix multiplications in
momentum and position representation, intervened by Fast
Fourier Transformations for the passage between both
representations.

In each propagation step, the kicking strength is determined
according to the following procedure. The regular kicking
strength is set to $K=7.5$. The noisy perturbations $k_n$ are
taken from a uniform box distribution over an interval
$(-W,W)$, so that $\kappa=W^2/3$. This noise is only applied at
times selected by a renewal process, which we generate by the
so-called Yule-Simon distributions  \cite{sim55}
\begin{equation}\label{eq:yule}
\omega (\tau) = \frac{\alpha \,\Gamma(\tau)\Gamma(\alpha
+1)}{\Gamma(\tau +\alpha+1)}.
\end{equation}
These distributions are paradigms of integer-valued probability
distributions with a power-law tail,
\begin{equation}\label{eq:yuleasym}
\omega(\tau)\sim \alpha\,
\Gamma(\alpha+1)\tau^{-1-\alpha}.
\end{equation}
The mean waiting time is given by
\begin{equation}
\overline\tau=\frac{\alpha}{\alpha-1}\quad(\alpha>1)
\end{equation}
and diverges for $\alpha<1$. In the latter case, the asymptotic
time dependence of the sprinkling distribution is
\begin{equation}
f(t)\sim \frac{\sin(\pi\alpha)}{\pi \Gamma(\alpha +1)}t^{\alpha-1}
.
\end{equation}

These properties directly follow from the Laplace transform of the
waiting-time distribution, which reads
\begin{equation}
\tilde
w(u)=\frac{\alpha}{\alpha+1}e^{-u}{}_2F_1(1,1,\alpha+2;e^{-u}),
\end{equation}
where ${}_2F_1$ is the hypergeometric function.

\begin{figure}
\epsfxsize=\columnwidth \epsffile{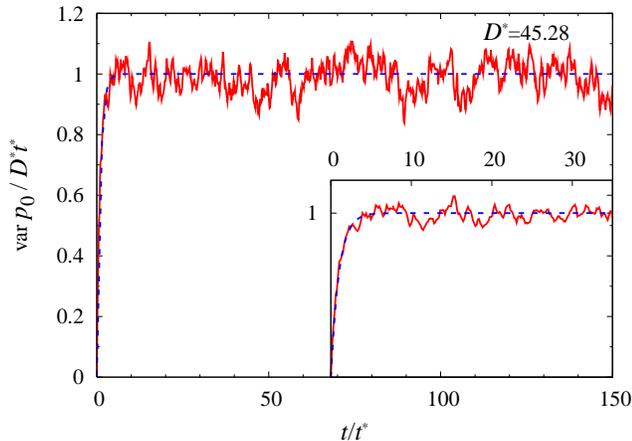}
\caption{\label{fig:2}(color online). Time dependence of the
variance of momentum $\var p_0(t)$ for a noiseless kicked rotator
with regular kicking strength $K=7.5$. The dashed curve is the
theoretical prediction  (\ref{eq:varp0}) of dynamical-localization
theory \cite{izr90}. A single-parameter fit delivers the value
$D^*=\hbar^2t^*=45.28$, which is used throughout the rest of the
present work. The inset zooms in onto the region where dynamical
localization is established.}
\end{figure}

The variance of the momentum for the kicked rotator in the absence
of noise is shown in Fig.\ \ref{fig:2}. A single-parameter fit to
the prediction (\ref{eq:varp0}) of dynamical-localization theory
delivers the value $D^*=\hbar^2t^*=45.28$, which is our only
fitting parameter and will not be adjusted any further throughout
the rest of this work.

\section{Decoherence in quasienergy space\label{sec:theory}}

The effect of noise on the quantum kicked rotator has been
first investigated by  Ott, Antonsen and Hanson, who developed
the following intuitive picture \cite{ott84}. In the absence of
noise the quantum system is fully coherent: the quasienergy
eigenstates of the Floquet operator $F(K)$ are exponentially
localized in momentum space, chaotic diffusion is strongly
suppressed and the momentum diffusion constant asymptotically
vanishes. The effect of the external noise is to couple the
quasienergy states and to induce transitions between them. As a
result, dynamical localization is extenuated and quantum
diffusion takes place.

A systematic approach for calculating the non-zero quantum
diffusion constant in the presence of stationary noise has been
developed by Cohen \cite{coh91}. It is based on an analysis of
the decay of a decoherence factor which is related to the
survival probability of the quasienergy states of the Floquet
operator. In Ref.\ \cite{coh91} this decoherence factor was
calculated perturbatively for short times (the emphasis was on
stationary noise with arbitrary correlations, including the
intricate correlations of zero-temperature noise). In the
following, we use a simple random-phase approximation to
calculate the decoherence factor beyond the perturbative
short-time regime. This approximation is valid for random
uncorrelated noise events and remains immediately applicable to
noise generated by a renewal process, including the case of
nonergodic, nonstationary L\'evy noise.

\subsection{Quasienergy transitions and decoherence factor}

The quasienergy eigenstates $|r\rangle$ of the noiseless system
are defined by the eigenvalue equation
\begin{equation}
F(K)|r\rangle=\exp(-i\varepsilon_r)|r\rangle,
\end{equation}
where $\varepsilon_r$ is the quasienergy. A noise-free
propagation step does not alter the quasienergy state, while a
noisy kick introduces transitions with amplitude
\begin{equation}\label{eq:transition}
\langle s | F(K_n)| s'\rangle=\exp(-i \varepsilon_{s}) \langle s |
\exp(-i\hbar^{-1}k_{n}\cos\theta)| s'\rangle,
\end{equation}
where $s'$ and $s$ are the quasienergy index before and after the
propagation step, respectively.

Over a larger time interval $[t'',t')$,  the survival amplitude in
a quasienergy state $|r\rangle$ is defined as
\begin{equation}\label{eq:surv}
A_r(t',t'')= \langle r|{\cal T}\prod_{n=t''}^{t'-1}
F(K_n)|r\rangle \exp[i(t'-t'')\varepsilon_r],
\end{equation}
where ${\cal T}$ is the time-ordering operator. In this expression,
we have explicitly provided for two time arguments, as is required
when working with a renewal process which may produce
nonstationary noise. The exponential factor compensates for the
dynamical accumulation of quasienergy phases in the noiseless
system, for which $A_r(t',t'')=1$.

The averaged survival amplitude is denoted by
\begin{equation}\label{eq:d1}
{\cal D}^{(1)}(t',t'') =\overline{A_r(t',t'')},
\end{equation}
where the average is over the index $r$ as well as over the timing
and amplitude of the noise.

The decoherence factor ${\cal D}(t',t'')$ that later appears
in the momentum spreading and in the decay of the purity is
given by
\begin{equation}\label{eq:d}
{\cal D}(t',t'') =\overline{A_r(t',t'')A_s^*(t',t'')},
\end{equation}
where the average now extends over both quasienergy indices $r\neq
s$.

\subsection{Average over amplitude noise}

In order to analyze the decoherence factor, we first assume that
the timing of noise events is fixed and average over the random
detunings $k_n$ of the noise events, as well as over the
quasienergy indices. These averages become straightforward within
a random-phase approximation, which uses that the random detunings
are uncorrelated and moreover exploits features of the complex
quantum dynamics encoded in the quasienergy transition amplitudes
(\ref{eq:transition}). For this we have to distinguish amplitudes
which preserve the quasienergy index (diagonal matrix elements,
which are of modulus one if there is no noise and less than one in
the presence of noise) from amplitudes describing transitions
between quasienergy states (off-diagonal matrix elements, which
are only finite in the presence of noise).

For noise strength $\kappa\ll\hbar^2$ the averaged modulus of the
diagonal matrix elements in Eq.\ (\ref{eq:transition}) can be
obtained in a simple expansion,
\begin{equation}
\overline {\exp(-i\hbar^{-1}k_{l}\cos\theta)}\approx
1-(\kappa/2\hbar^2)\,\overline{\cos^2\theta}=1-\kappa/4\hbar^2,
\end{equation}
which delivers the survival time
\begin{equation} t_c\simeq 2\hbar^2/\kappa.
\end{equation}
We will shortly see that $t_c$ coincides with the decoherence
time for conventional amplitude noise, which acts stationarily
in every kick. The condition set for the noise strength implies
$t_c\gg 1$. This assumption is not very restrictive, since it
only prevents total decoherence by a single event, but does not
impose any restriction on $t_c$ with respect to the
localization time $t^*\simeq K^2/2\hbar^2$, which is also much
larger than unity. We hence reserve the notion of {\em weak
noise} for the more restrictive condition $\kappa\ll
\hbar^4/K^2$, which implies $t_c\gg t^*$.

Each noise event mixes the initial quasienergy state into about
$\xi=t^*/2\gg 1$ states whose localization centers are in reach
of the localization length \cite{shep}. Each individual
off-diagonal transition amplitude hence is small, and for weak
noise ($t_c\gg t^*$) the averaged off-diagonal matrix elements
indeed can be trivially neglected. The averaged Floquet
operator $F(K_n)$ is then simply approximated by
\begin{equation} \langle s | \overline {F(K_n)}| s'\rangle= \delta_{s,s'}\mathrm{
diag}[e^{-i\varepsilon_s-1/(2t_{c})}], \label{eq:rpa}
\end{equation}
which is diagonal in quasienergy space.

When $t_c$ and $t^*$ are of comparable order, individual
contributions which include quasienergy transitions are still
small, but in the composed survival amplitude (\ref{eq:surv}) one
cannot disregard the proliferation of such terms when the number
of transitions is increased. It then becomes necessary to exploit
the consequences of the complex dynamics of the kicked rotator,
which induces random phases into the off-diagonal transition
matrix elements. The random phase has several separate origins:

(1) For fixed $s$ and $s'$, the transition amplitude
(\ref{eq:transition}) sensitively depends on the random
detuning $k_n$ of the kick.

(2) In momentum space, the quasienergy states have an
exponentially decaying envelope, but a quasi-random internal
structure. Consequently, even after the average over the detuning,
the phase of the transition amplitude depends strongly on the
quasienergy indices $s$ and $s'$.

In the ensuing random-phase approximation, the phase of the
off-diagonal transition amplitudes is assumed to be uniformly
distributed in $[0,2\pi)$, while the squared modulus of these
matrix elements is of order $2/(t_c t^*)$ (corresponding to the
bandwidth $\xi=t^*/2$). The squared modulus of the diagonal
elements is given by $\exp(-1/t_c)\approx 1-1/t_c$.

In this approximation, the averaged Floquet operator $F(K_n)$
simply retains its diagonal form (\ref{eq:rpa}). This is all we
need for the computation of the averaged survival amplitude
${\cal D}^{(1)}(t',t'')$ and leads to the intermediate
expression
\begin{equation}
\label{eq:d1intermed}
 {\cal D}^{(1)}(t',t'') =\overline{\exp[-{\cal N}(t',t'')/(2t_{c})]},
\end{equation}
where ${\cal N}(t',t'')$ is the inverse random time of the renewal
process as introduced in Sec. \ref{sec:model}.

The random phase approximation also establishes a direct
connection between ${\cal D}^{(1)}(t',t'')$ and the decoherence
factor ${\cal D}(t',t'')$. Since in  Eq.\ (\ref{eq:d}) the two
quasienergy indices $r$ and $s$ are not identical, there are no
contributions which link the quasienergy states in exactly the
same sequence. The decoherence factor therefore immediately
factorizes into the averaged survival amplitudes of the two
quasienergy states, yielding
\begin{equation}
\label{eq:dintermed} {\cal D}(t',t'') =\overline{[{\cal
D}^{(1)}(t',t'')]^2}=\overline{\exp[-{\cal N}(t',t'')/t_{c}]}.
\end{equation}
The remaining average over the timing of the noise events does not
factorize, but because of the simple exponential form of Eq.\
(\ref{eq:d1intermed}), merely amounts to a factor of two in the
exponent.  We now analyze this average for different types of
renewal processes.

\subsection{Average over timing noise}

In the conventional case of stationary noise with
$\omega(\tau)=\delta_{\tau,1}$, where timing noise is absent, the
inverse random time, ${\cal N}(t',t'')=t''-t'$, does not fluctuate.
According to Eq.\ (\ref{eq:dintermed}), the decoherence factor
\begin{equation}
\label{eq:d1stat}
 {\cal D}(t',t'') = \exp[-(t'-t'')/t_{c}] \quad \mbox{(stationary
 noise)}
\end{equation}
then decays exponentially over time, and $t_c$ is identical to the
coherence time.

\begin{figure*}[t]
\epsfxsize=.8\textwidth \epsffile{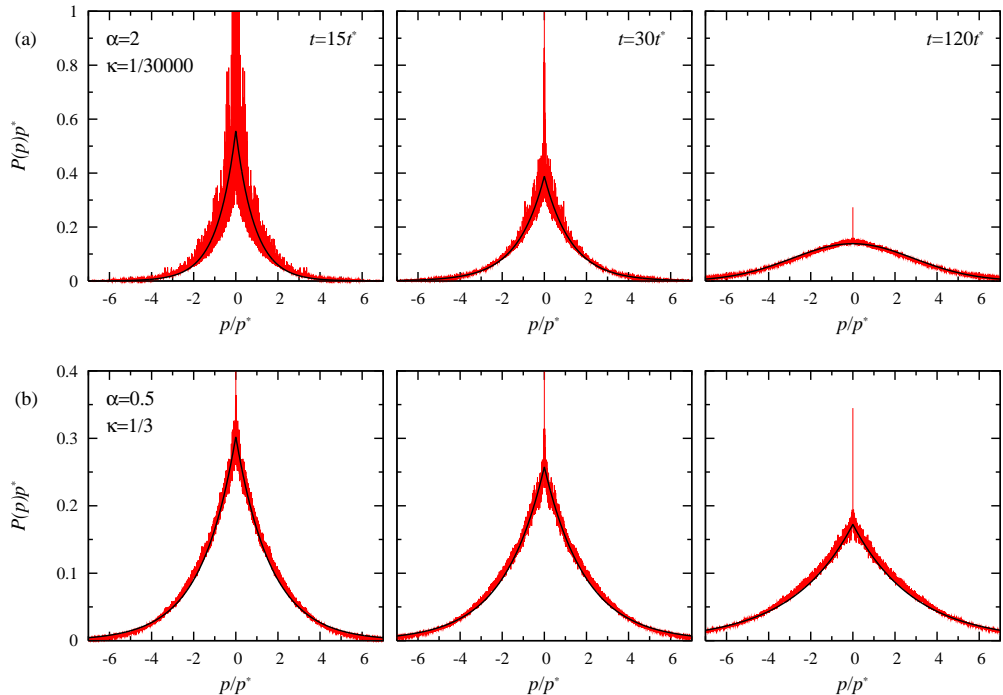} \caption{(color
online). Averaged distribution of momenta $P(p;t)$ at selected
times $t$ for kicked rotators which are subjected to L{\'e}vy
noise generated by a Yule-Simon distribution. The average is over
$10^3$ realizations. The smooth curves are fits to double-sided
exponential and Gaussian distributions. The unit of momentum is
the localization length $p^*=D^*/\hbar$. (a) In the upper panels
$\alpha=2.0$, $\kappa=1/30000$, corresponding to a decoherence
time $t_c=12.3\,t^*$ (where $t^*$ is the localization time). At
$t=15\,t^*\approx 1.2\,t_c$ the distribution function fluctuates
around a double-sided exponential function. At $t=30\,t^*\approx
2.4\,t_c$ the distribution function is still exponential, but the
fluctuations are suppressed. At $t=120\,t^*\approx 9.8\,t_c$ the
distribution function is well described by a Gaussian. (b) In the
lower panels $\alpha=0.5$, $\kappa=1/3$, corresponding to stronger
but nonstationary, increasingly rare noise events. In this case, a
decoherence time cannot be defined and the distributions function
remains exponential for all times $t>t^*$. Note that the peak at
$p=0$ arises from coherent backscattering. \label{fig:3}}
\end{figure*}

For noise generated by a renewal process with power-law
waiting-time distributions, the subsequent analysis depends on
whether the process is characterized by a mean waiting time
$\overline \tau$, or shows the
 nonstationary
features of L\'evy noise. This bifurcation can be made explicit by
realizing that Eq.\ (\ref{eq:d1intermed}) is identical to the
moment-generating function ${\cal M}(z;t',t'')$ of the number of
noise events, introduced in Eq.\ (\ref{eq:mdef}),
 which has to be evaluated at the noise-strength
dependent value $z=-1/t_c$. We therefore can immediately exploit the
results of Sec.\ \ref{sec:levy}. (Since we assume $t_c\gg 1$ we
can set $e^z-1 \approx -1/t_c$.)

It follows from Eq.\ (\ref{eq:statdecay}) that, when the noise is
asymptotically stationary ($\alpha>1$), the decoherence factor
retains its exponential form,
\begin{equation} \label{eq:d1agtr1}
 {\cal D}(t',t'') \sim \exp[-(t'-t'')/\overline{\tau} t_{c}] \quad
 (\alpha>1).
\end{equation}
The effective coherence time $\overline\tau t_{c}$ is directly
proportional to the mean waiting time of the noisy kicks and
therefore increases with decreasing exponent $\alpha$. In
particular, the effective coherence time becomes {\it infinitely}
large when $\alpha$ drops below unity. According to Eq.\
(\ref{eq:ma}), the functional form of the decoherence factor then
changes to
\begin{equation}
{\cal D}(t,0)\approx E_\alpha\left[-
\Gamma(1+\alpha)\overline{{\cal N}(t,0)}/t_c
\right] \quad (\alpha<1). \label{eq:MLp}
\end{equation}

After a (typically short) transient, $\overline{{\cal N}(t,0)}\sim
t^{\alpha}\sin(\pi\alpha)/(\pi c)$. Hence, we have the remarkable
result that L{\'e}vy noise induces  a nonexponential loss of
coherence which typically starts out as a stretched exponential,
\begin{equation}
{\cal D}(t,0) \simeq \exp[t^\alpha /(\Gamma(-\alpha) c
t_c)],\quad\mbox{(initial decay)},
\end{equation}
and for large times crosses over to a power law,
\begin{equation}\label{eq:asym}
{\cal D}(t,0)\simeq (c t_c/\alpha) t^{-\alpha}
\quad\mbox{(asymptotic decay)}.
\end{equation}
The functional dependence of the decoherence factor can thus be
changed in a controlled manner by tuning the value of the exponent
$\alpha$. We now explore how this affects the observable
properties of the kicked rotator.

\begin{figure}[t]
\epsfxsize=\columnwidth \epsffile{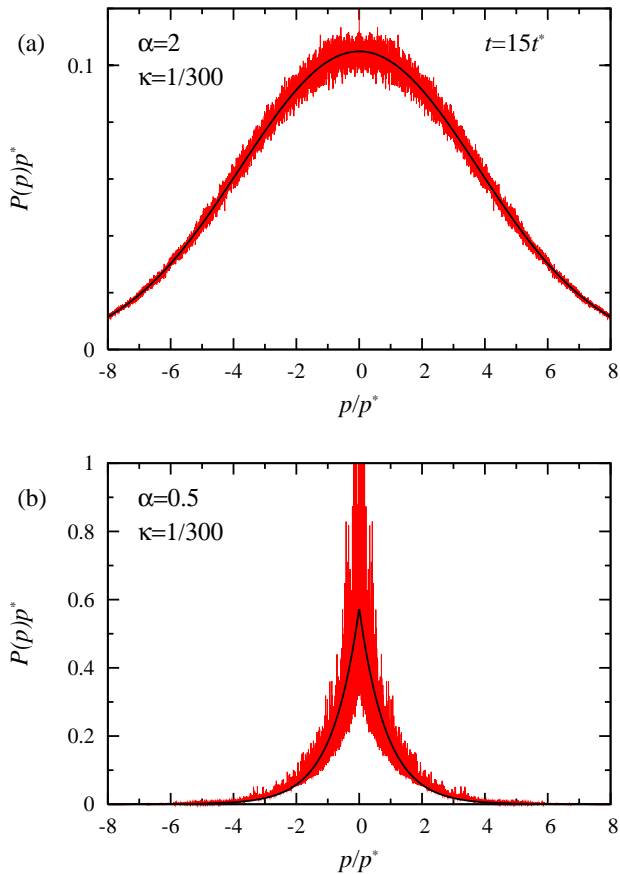} \caption{(color
online). Same as the left panels in Fig.\ \ref{fig:3}, but for
$\kappa=1/300$ ($t_c=0.12\,t^*$), so that $t=15\,t^*\gg t_c$.
\label{fig:4}}
\end{figure}

\section{Momentum spreading\label{sec:momentum}}

In the atom-optical experiments, the momentum can be observed by a
time-of-flight measurement, where the kicking sequence is
terminated at a specified time $t$ and the atom cloud is allowed
to expand ballistically to a much larger size. A snapshot is then
taken and the travel distance is converted into the momentum. This
technique provides direct access to the complete momentum
distribution $P(p;t)$.

In the presence of dynamical localization, the momentum
distribution shows the exponential envelope of the atomic
wavefunctions in momentum space, while a Gaussian envelope is
expected when diffusive momentum spreading is induced by
stationary timing or amplitude noise. In order to give an early
indication of the effects of nonexponential decoherence, we
compare in Fig.\ \ref{fig:3} the momentum distribution of L\'evy
kicked rotators with asymptotically stationary and nonstationary
noise ($\alpha=2$ and $\alpha=1/2$, respectively). For $\alpha =
2$, the strength of the noise is set to  $\kappa=1/30000$,
corresponding to a decoherence time $t_c=12.3\,t^*$. For
$t\lesssim t_c$ the distribution function is well described by a
double-sided exponential function; when the time is increased it
eventually evolves into a Gaussian. This follows the expectations
for conventional stationary noise. By contrast, for $\alpha=1/2$,
where a well defined coherence time does not exist, the
probability density keeps its double-sided exponential shape at
all times. This is the case even though we increased the strength
of the individual noise events by four orders of magnitude to
$\kappa=1/3$, so that $t_c=0.00123\,t^*$ is much reduced. This
indicates that diffusive momentum spreading is never attained,
even in the limit of very strong noise events.

Figure \ref{fig:4} shows the momentum distribution at $t=15t^*$
for the more moderate noise strength $\kappa=1/300$
($t_c=0.12\,t^*$) which is used in several subsequent figures in
this work. For $\alpha=2.0$, the momentum distribution
 is always well-approximated by a Gaussian, while for
 $\alpha=0.5$ and $t>t^*$,
it is well-approximated by a double-sided exponential function
as soon as $t>t^*$.

We now proceed to characterize the evolution of the momentum
distribution by two quantities, the variance $\var p$ and the
inverse participation ratio (IPR).

\subsection{Variance of momentum}

The temporal evolution of the momentum distribution $P(p;t)$ has
two origins. The first component is the spreading of the momentum
wavefunction of an individual atom with respect to the mean
momentum of the atom. This momentum spreading  is characterized by the variance,
\begin{equation}\label{eq:vardef}
\var p(t)=\overline{\langle p^2\rangle-\langle p\rangle^2}.
\end{equation}
As indicated, this variance is computed from the
quantum-mechanical expectation values, and then averaged over
the noise. The second component of the evolution of the momentum distribution is the spreading of the mean
momenta of the different atoms. Hence, in general, the variance
(\ref{eq:vardef}) is smaller than the total variance computed
from the momentum distribution $P(p;t)$. In the particular case
of the kicked rotator, however, this difference becomes
negligible, due to the symmetry $(\theta,p)\to (-\theta,-p)$
which holds even in presence of the noise.

In the following, we first establish a general relation between
the variance of the momentum of the noise-free and noisy systems,
and then evaluate this relation for the L{\'e}vy kicked rotator.
We will then find that nonexponential decoherence manifests itself
in a subdiffusive spreading of the momentum.

As in the earlier theories of delocalization due to decoherence
\cite{coh91}, the starting point of our considerations is the
force-force correlation function,
\begin{equation}
C(t',t'')=\overline{\langle
K_{t'}K_{t''}\sin\theta_{t'}\sin\theta_{t''}\rangle},
\end{equation}
which is composed of the momentum increments in an individual time
step, as given by Eq. (\ref{eq:pmap}). In the above equation, $\theta_t$ is the angle operator
in the Heisenberg picture. The average includes an average over
the initial momentum of the particle, so that the expectation
value is replaced by the normalized trace $N^{-1}\tr_N$, where $N$
is the Hilbert space dimension.

 The total change of the momentum is
obtained by integrating the force over time. The
variance therefore reads
\begin{equation}\label{eq:varpc} \var p(t)=\sum_{t',t''=0}^{t-1}
C(t',t'').
\end{equation}

In order to establish the relation to the decoherence factor,
the force-force correlation function is expanded in the basis
of quasienergy eigenstates \cite{coh91},
\begin{widetext}
\begin{eqnarray}
&&C(t',t'')=\overline{\sum_{\{r_n,s_n\}}^\prime \langle
 r_{t'} | \sin\theta| s_{t'}  \rangle  \langle
s_{t''} | \sin\theta| r_{t''}  \rangle
K_{t'}K_{t''}\prod_{l,m=t''}^{t'-1} \langle r_l| F (K_l)|
r_{l+1}\rangle^* \langle s_{m+1}| F (K_m)| s_m\rangle}.
\label{eq:c}
\end{eqnarray}
\end{widetext}
The prime in the sum over quasienergy indices excludes the initial
index $r_{t''}$, which originates from the normalized trace and
hence instead is averaged over.

We first apply the random-phase approximation to the matrix
elements of the sine kicking potential, which in principle selects
three types of terms: (i) terms with $r_{t'}=r_{t''}$,
$s_{t'}=s_{t''}$, but $r_{t'}\neq s_{t'}$; (ii) terms with
$r_{t'}=s_{t'}$, $r_{t''}=s_{t''}$, but $r_{t'}\neq r_{t''}$;
(iii) terms with $r_{t'}=r_{t''}=s_{t'}=s_{t''}$.

The terms of type (ii) and (iii) involve the expectation value of
the force, which vanishes in the specific case of the kicked
rotator, thanks to the symmetry $(\theta,p)\to (-\theta,-p)$ (a
careful analysis of such terms will be required for the purity,
see Section \ref{sec:purity}). For the terms of type (i), on the
other hand, the product of transition matrix elements $ \langle
r_l| F (K_l)| r_{l+1}\rangle$  in quasienergy space exactly
arranges itself into the definition of the decoherence factor
(\ref{eq:d}). This correspondence also holds when $C(t',t'')$ is
evaluated for $t'<t''$, provided we define ${\cal D}(t',t'')={\cal
D}(t'',t')$.

For $t'=t''$, the average over the noise also gives rise to a
contribution to the typical force acting on the particle of the
form $\overline{K_{t'}^2}\langle \sin\theta^2\rangle=K^2/2
+(\kappa/2) f(t')$, where $f(t)$ is the sprinkling distribution.
Collecting all terms, we arrive at
\begin{equation}\label{eq:cres}
C(t',t'')=C_0(t',t''){\cal
D}(t',t'')+\frac{\kappa}{2}f(t')\delta_{t',t''},
\end{equation}
where  $C_0(t',t'')$ is the force-force correlation function in
the absence of  noise. Besides the correction from the sprinkling
distribution, which is small for weak noise, this recovers the
result from the perturbative short-time theory in Ref.\
\cite{coh91}. It follows that the correct extrapolation of the
decoherence factor beyond the initial decay is given by Eq.\
(\ref{eq:dintermed}).

The variance of the momentum follows when Eq.\ (\ref{eq:cres})
is inserted into  Eq.\ (\ref{eq:varpc}). It is then convenient
to pass to a continuous time argument and perform integrations
by parts, so that all force correlations are expressed in terms
of the variance of the momentum in absence of the noise, which
we denote by $\var p_0(t)$. We then obtain as one of our main
results the following expression,
\begin{eqnarray}
\label{eq:varp} &&\var p(t)=\var p_0(t)\,{\cal
D}(t,0)+\frac{\kappa}{2}\overline{{\cal N}(t,0)}\nonumber\\
&& +\int_0^t ds\,\var p_0(t-s)\,\partial_s {\cal
D}(t,s)\nonumber\\
&& -\int_0^t ds\,\var p_0(s)\,\partial_s {\cal D}(s,0)
\nonumber\\
&&-\int_0^t ds'\int_0^{s'} ds''\,\var p_0(s'-s'')\,\partial_{s'}
\partial_{s''}{\cal D}(s',s'')  .\quad
\end{eqnarray}
The second term on the right hand side can be neglected for small
noise strength $\kappa$. Equation (\ref{eq:varp})  expresses
the momentum variance in presence of noise exclusively in terms of
the variance in absence of the noise, and derivatives of the
decoherence factor ${\cal D}(t',t'')$, which play the role of
memory kernels. Because of the monotonicity properties of ${\cal
D}(t',t'')$ all terms are positive. Since the result is expressed
directly in terms of $\var p_0$, Eq.\ (\ref{eq:varp}) moreover
accounts for all dynamical correlations of the noiseless dynamics.
In particular, the random-phase approximation used in the
derivation only relies on the complex quantum-dynamics of the
noiseless system, but does not require that it displays dynamical
localization.

In exploring the consequences of Eq.\ (\ref{eq:varp}) for the
kicked rotator, we combine Eq.\ (\ref{eq:varp0}) for the momentum
variance in absence of the noise with the various decoherence
factors found in Section \ref{sec:theory}. First, let us again
make contact with the known results for stationary noise, for
which the decoherence factor decays exponentially as given in Eq.\
(\ref{eq:d1stat}). The corresponding momentum spreading is
\begin{equation}
\var p= \frac{D^*}{1+t_c/t^*}t+
\frac{D^*t^*}{(1+t^*/t_c)^2}[1-\exp(-t/t^*-t/t_c)].
\label{eq:crossover}
\end{equation}
 This result was first proposed in Ref.\
\cite{coh91} as one of two possible extrapolations of the
perturbative short-time results, and has been successfully applied
to quantify the decoherence process in the experiment of Ref.\
\cite{kla98}. The present derivation shows that the other proposed
extrapolation indeed was rightfully discarded. For large times,
the second term approaches a constant and can be neglected in
comparison to the first term, which increases linearly in time.
Hence, for stationary noise when quantum coherence is lost
exponentially, the momentum diffuses asymptotically with a
renormalized diffusion constant $D^*/(1+t_c/t^*)$.

In the presence of asymptotically stationary noise generated by
a renewal process with $\alpha>1$, the decoherence function is
given by Eq.\ (\ref{eq:d1agtr1}), in which the coherence time
is modified by the mean waiting $\overline\tau$. This does not
induce any qualitative change in the momentum spreading, which
is still diffusive according to a renormalized diffusion
constant $D^*/(1+\overline\tau t_c/t^*)$.

As $\alpha$ approaches unity from above, the decoherence time
increases, and the quantum diffusion constant vanishes at
$\alpha=1$. It is clear that this has to be accompanied by a
qualitative change of the momentum spreading itself. For
$\alpha<1$, the asymptotic spreading is dominated by the last
term in Eq.\ (\ref{eq:varp}), which can be approximated as
\begin{eqnarray}
\label{eq:varpas} \var p(t)&\sim& -\var p_0(\infty)\int_0^t
ds'\int_0^{s'} ds''\,\,\partial_{s'}
\partial_{s''}{\cal D}(s',s'')
\nonumber \\
&=&\frac{D^*t^*}{t_c}\,\overline{{\cal N}(t,0)}.
\end{eqnarray}
Using the power-law asymptotics (\ref{eq:nalphaleq1}) of the
inverse random time for $\alpha\leq 1$, we immediately find that the momentum spreads {\em subdiffusively},
\begin{equation}
\var p(t) 
\sim \frac{D^*t^*}{t_c}\frac{\sin\pi\alpha}{\pi c} t^{\alpha}.
\label{eq:subdiff}
\end{equation}

\begin{center}
\begin{figure}[t]
\epsfxsize=\columnwidth \epsffile{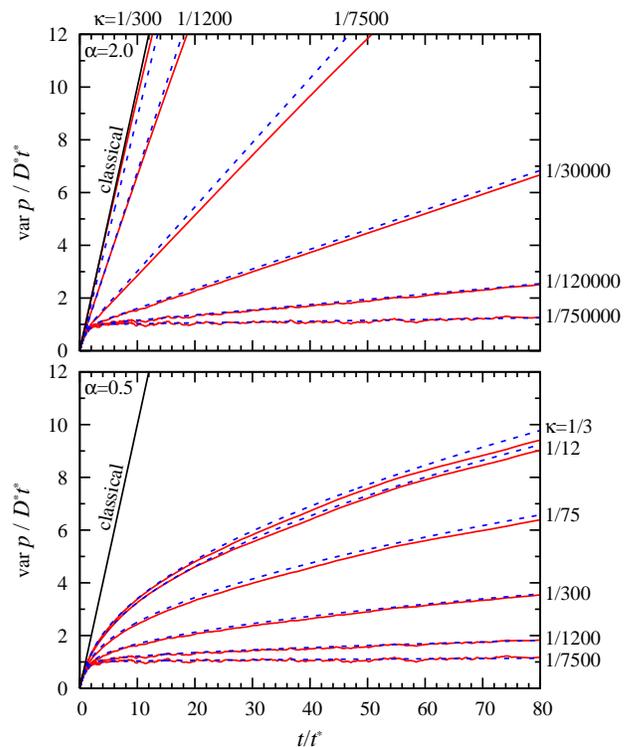} \caption{(color
online). Time dependence of the momentum spreading $\var p(t)$ for
kicked rotators which are subjected to L{\'e}vy noise generated by
a Yule-Simon distribution with noise exponent $\alpha=2.0$ (panel
a) and  $\alpha=0.5$ (panel b). The results of numerical
computations (solid curves) are compared to the theoretical
prediction from Eq.\ (\ref{eq:varp}) (dashed curves). The straight
solid line labelled "classical" is the classical diffusion.
\label{fig:5}}
\end{figure}
\end{center}

The full time dependence of the momentum variance of L\'evy kicked
rotators with asymptotically stationary or non-stationary noise is
shown in Fig.\ \ref{fig:5}.  The upper panel shows the results for
$\alpha=2$, for which the momentum spreading is diffusive. The
classical diffusion is attained for moderate noise strength
$\kappa\gtrsim 1/300$. The lower panel shows the results for
$\alpha=1/2$, and confirms that in this case the spreading is
subdiffusive, $\var p \propto t^{1/2}$. Note that this behavior
persists even when the strength of the individual noise events is
large. The theoretical predictions in the plots follow from Eq.\
(\ref{eq:varp}), where $\var p_0(t)$ is taken from Eq.\
(\ref{eq:varp0}), while ${\cal D}(t',t'')$ is determined for Eq.\
(\ref{eq:dintermed}) [i.e., the moment-generating function of
${\cal N}(t',t'')$ for the Yule-Simon distribution is evaluated at
$z=-1/t_c=-\kappa/2\hbar^2$]. There is good agreement between the
numerical results and the theoretical predictions.

\subsection{Inverse participation ratio\label{sec:ipr}}

In this section we study an alternative measure of the momentum
spreading, the inverse participation ratio (IPR), and show that
this quantity also reveals  the existence of two distinct
(exponential and non-exponential)  decoherence regimes. The
inverse participation ratio is defined as
\begin{equation}
\label{eq:purity2} \mathrm{IPR}(t)=
\sum_{n}\overline{|\psi_{n}(t)|^4},
\end{equation}
where $\psi_n$ are the amplitudes of the wavefunction in a
fixed basis. The IPR can be interpreted as a measure for the
inverse dimension of an effective Hilbert space explored by the
quantum dynamics. In general, the IPR is basis dependent. In
the following, we provide a quasiclassical estimate of the IPR
by introducing basis states which cover Planck cells of area
$2\pi \hbar$ in phase-space and are centered at phase-space
positions $(\theta_n,p_n)$. This estimate turns out to be
independent on the details of the quasiclassical basis.

For a given time, the complex amplitudes $\psi_{n}$ fluctuate as
$n$ is varied, corresponding to speckle noise in the phase-space
position $(\theta_n,p_n)$. In order to quantify these fluctuations,
we introduce a phase-space function $P(\theta_n,p_n;t)$  such that
the statistical average of the squared amplitude is
\begin{equation}
\label{eqi1} \overline{|\psi_{n}(t)|^2} \equiv 2\pi\hbar
P(\theta_n,p_n;t).
\end{equation}
By assuming that the real and imaginary parts of $\psi_{n}$ are
Gaussian  distributed and independent, we  have
\begin{equation}
\label{eqi2} \overline{ |\psi_{n}(t)|^4} =2
\overline{|\psi_{n}(t)|^2}^2.
\end{equation}
In the quasiclassical approximation, the function
$P(\theta,p;t)$ is now identified with a smooth, properly normalized phase space
density with continuous variables  $\theta$ and $p$. By combining Eqs.~(\ref{eqi1})
 and (\ref{eqi2}), the  IPR can then be written in the form
\begin{eqnarray}
\label{eq:purity3} \mbox{ IPR}(t)= \int
\frac{d\theta\,dp}{2\pi\hbar}\times 2 [2\pi\hbar P(\theta,p;t)]^2\
.
\end{eqnarray}
The above expression is invariant under canonical
transformations of the phase-space variables and hence exhibits
the advertised independence of details of the quasi-classical
basis.

In the L\'evy kicked rotator, the quasiclassical probability
density becomes $\theta$-independent after a short time, so that
to a good approximation $P(\theta,p;t)\simeq P(p;t)/2\pi$. This
leads to a compact expression for the IPR,
\begin{equation}
\label{eqi3} \mbox{IPR}(t) =2\hbar\int dp\, P^2(p;t).
\end{equation}

\begin{figure}[t]
\epsfxsize=\columnwidth \epsffile{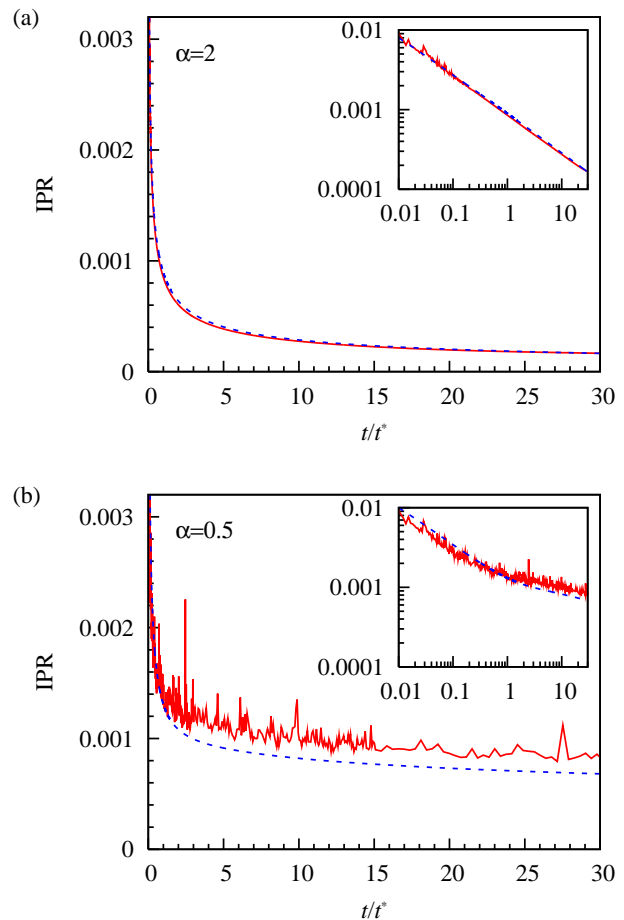} \caption{(color
online). Time dependence of the inverse participation ratio IPR
for kicked rotators which are subjected to L{\'e}vy noise
generated by a Yule-Simon distribution with $\kappa=1/300$. The
main panels are single-logarithmic while the insets show the same
data in a double-logarithmic presentation. (a) Noise with exponent
$\alpha=2.0$, corresponding to a decoherence time $t_c=0.12\,t^*$
(where $t^*$ is the localization time) so that the momentum
distribution $P(p;t)$ is always well-approximated by a Gaussian.
The dashed line is the quasi-classical prediction
(\ref{eq:purity4a}).
 (b)  Noise with exponent $\alpha=0.5$,
where for $t>t^*$ the momentum distribution $P(p;t)$ is always
well-approximated by a double-sided exponential function. The
dashed line is the quasi-classical prediction (\ref{eq:purity4b}).
 \label{fig:6}}
\end{figure}

The integral in Eq.~(\ref{eqi3}) can be calculated in two simple
cases: For a Gaussian profile, $P(p;t)=(2\pi \var
p(t))^{-1/2}\exp[-p^2/2\var p(t)]$, typical for diffusive
spreading, we obtain
\begin{equation}
\label{eq:purity4a} \mbox{IPR}(t)= \frac{\hbar}{\sqrt{\pi\var
p(t)}} \quad\mbox{(Gaussian profile)}.
\end{equation}
On the other hand, for a double-sided exponential probability distribution,
$P(p;t)=(\lambda/2)\exp(-\lambda|p|)$ with $\lambda=\sqrt{2/\var
p}$, we find
\begin{equation}
\label{eq:purity4b} \mbox{IPR}(t)= \frac{\hbar}{\sqrt{2\var p(t)}}
\quad\mbox{(exponential profile)}.
\end{equation}
Interestingly, in  both Eqs.~(\ref{eq:purity4a}) and
(\ref{eq:purity4b}) the IPR is proportional to $\hbar/\sqrt{\var
p(t)}$. In view of Eq.~(\ref{eq:subdiff}), we hence expect that
the asymptotic decay of the IPR is  a power-law with exponent
$-1/2$ for  $\alpha>1$, while for $\alpha<1$ the exponent is
reduced to $-\alpha/2$. These predictions are confirmed by the
numerical results shown in Fig.\ \ref{fig:6}. We can therefore
conclude that the quasiclassical expressions (\ref{eq:purity4a})
and (\ref{eq:purity4b}) offer a good quantitative estimate of the
IPR of the L\'evy kicked rotator.

\section{Purity and fidelity\label{sec:purity}}

A convenient quantity to further elucidate the loss of coherence
in a quantum-dynamical system is the fidelity
\cite{jalabert,jacquod,silvestrov,echo},
\begin{equation} {\cal
F}(t) = |\langle \psi(t)|\psi'(t)\rangle|^2,
\end{equation} where the states $\psi(t)$ and $\psi'(t)$ start out from the same initial condition
$\psi(0)=\psi'(0)$,  but are propagated with different
realizations of the noise.

A remarkable property of the fidelity is that it can be directly
probed in an echo experiment \cite{scha05}.
 In the
latter, the quantum state is first let to evolve up to a certain
time $t$ at which the dynamics is reversed, e.g., by applying a
pulse that inverts the momentum ($p\to-p$). When at time $2t$ the
momentum-reverting pulse is applied again, the system
approximately returns into its initial state. The fidelity
characterizes the quality of this echo, which is never perfect
because of imperfections in the momentum-reverting pulse and the
noise in the dynamics. The destruction of phase coherence in a
quantum system such as the kicked rotator can therefore be
directly studied in  echo-like experiments.

In the present section, we study various statistical aspects of the
fidelity  for the case that its decay is induced by L\'evy
noise.

\subsection{Purity}

The statistical average of the fidelity
\begin{equation} \label{eq:purdef}
\overline{{\cal F}(t)}=\tr \rho^2(t)
\end{equation}
is formally equivalent to the so-called purity of the statistical
mixture of the perturbed states, which is described by the density
operator
\begin{equation}
\rho(t)=\overline{|\psi(t)\rangle\langle\psi(t)|}.
\end{equation} The average is over the amplitude and
timing noise generated by the renewal process. The initial
state is fixed, so that the purity starts at $\tr
\rho^2(0)=1$. For completely incoherent superpositions the
purity tends to the value $1/N$, where $N$ is the Hilbert space
dimension.

In the following, we show that the initial and
intermediate decay of the purity is directly related to the
decoherence function ${\cal D}(t)$. For the long-time
asymptotics, we moreover establish a connection to the inverse
participation ratio.

 In the quasienergy
basis, starting from  Eq.\ (\ref{eq:purdef}), the purity can be
expanded as
\begin{widetext}
\begin{eqnarray}
&&\tr \rho^2(t)=
\overline{\!\!\!\!\!\sum'_{\{r_n,s_n,r'_n,s'_n\}} \!\!\!\!\!\psi_{r_0}
\psi^*_{r'_0} \psi^*_{s_0} \psi_{s'_0}\!\!\!\!\!\prod_{l,m,l',m'=0}^{t-1}\!\!\!\!\!
\langle r_{l+1} |F(K_l)| r_l\rangle  \langle r'_{l'+1}
|F(K'_{l'})| r'_{l'}\rangle ^* \langle s_{m+1} |F(K_m)| s_m\rangle
^*  \langle
s'_{m'+1} |F(K'_{m'})| s'_{m'}\rangle},\nonumber\\
\end{eqnarray}
\end{widetext}
where the prime in the sum denotes the constraint $r_t=r'_t$,
$s_t=s'_t$. Furthermore,  $\psi_r$ is the initial wavefunction,
expanded in the quasienergy eigenbasis, and the values $K_l$ and
$K_l'$ are obtained from independent realizations of the renewal
process, which differ both in the timing as well as in the
detunings of the noisy kicks.

We first apply the random-phase approximation to the expansion
coefficients $\psi_r$, which selects three groups of terms: (i)
terms with $r_0=r_0'$, $s_0=s_0'$, but $r_0\neq s_0$; (ii) terms
with $r_0=s_0$, $r_0'=s_0'$, but $r_0\neq r_0'$; (iii) terms with
$r_0=r_0'=s_0=s_0'$.

For each class of terms, the random-phase averages over the
detunings and timing of the noisy kicks can be carried out
following the procedure which we applied to the decoherence factor
and the survival probability in quasienergy space, described in
Section \ref{sec:theory}. For the terms of group (i), this
procedure selects only diagonal matrix elements with fixed
${r_n=r_n'\equiv r}$, ${s_n=s_n'\equiv s}$, just as encountered
for the decoherence factor. Since we now deal with two independent
renewal processes, they deliver a factor $\exp[-{\cal
N}(t,0)/t_c+{\cal N}'(t,0)/t_c]$ which depends on the sum of
inverse random times in the propagation of the two states. After
averaging over the timing noise of the two independent renewal
processes, these terms therefore deliver a factor ${\cal
D}^2(t,0)$.

Among the terms of group (ii), the random-phase approximation
requires to pair all amplitudes to probabilities, $r_n=s_n$,
$r_n'=s_n'$. In the calculation of the variance $\var p(t)$, these terms
could be neglected because they were multiplied by the vanishing
expectation value of the detuned force.  We shall denote these
contributions by ${\cal S}_{r\to s,\rm cl}(t,0)$.

Among the terms of group (iii), both terms which are completely
diagonal (type i) as well as terms which are paired to
probabilities  (type ii) survive. A neat separation of these terms
follows when we require that for $r=s$, the terms are called of
type (ii) when they contain at least one quasienergy change.
Appropriately attributing these terms to the groups (i) and (ii),
the purity then takes the form
\begin{equation}
\tr \rho^2(t)=\sum_{r,s} |\psi_r|^2|\psi_s|^2 [{\cal
D}^2(t,0)+{\cal S}_{r\to s,\rm cl}(t,0)],
\end{equation}
where the restriction $r\neq s$ is now lifted.

For the term proportional to ${\cal D}^2(t,0)$ we can now exploit
the normalization of the initial wavefunction. Looking at the
second term, it easy to see that these contributions can be
neglected for short times $t\ll t_c$, since per definition they
contain at least one quasi-energy transition. The significance of
these terms  becomes apparent when one re-evaluates the purity for
long times. For this we start from the general expression,
\begin{equation}
\tr \rho^2(t) =
\sum_{nm}\overline{\psi_{n}^*(t)\psi_{n}'(t)\psi_{m}^{\prime
*}(t)\psi_{m}(t)},
\end{equation}
and adopt the quasi-classical basis in which we computed the
IPR (see Section \ref{sec:ipr}). For large times $t\gg t_c$ it
is reasonable to assume that the relative phases of the
wavefunction expansion coefficients $\psi_{n}$ and $\psi_{n}'$
in this basis are completely randomized. Assuming furthermore
$|\psi_{n}|^2 \approx|\psi_{n}'|^2$, we find that $\tr
\rho^2(t)$ asymptotically approaches the  IPR as defined in
Eq.~(\ref{eq:purity2}). The terms of group (ii) are therefore
associated to the eventual saturation of the purity at the
inverse of the effective Hilbert space dimension. We hence
arrive at our main result for the purity,
\begin{equation}\label{eq:purity}
\tr \rho^2(t)\approx{\cal D}^2(t,0)+{\rm IPR}(t).
\end{equation}

The initial and intermediate decay of the purity is given by the
first term, which is simply the square of the decoherence function
${\cal D}(t,0)$. For stationary noise the initial decay is thus
exponential, $\tr \rho^2(t)=\exp(-2t/t_c)$. A similar behavior is
predicted for L{\'e}vy noise with $\alpha>1$, for which $t_c$
replaced by $\overline\tau t_c$. For nonstationary noise with
$\alpha<1$, Eq.\ (\ref{eq:MLp}) predicts that the initial and
intermediate decay shows the functional features of the squared
Mittag-Leffler function, i.e., a stretched exponential which for
later times crosses over to a power law with exponent $-2\alpha$.

The long time behavior of the purity is given by the saturation at
the inverse effective Hilbert space dimension, which we estimate
in terms of the inverse participation ratio. According to Eqs.\
(\ref{eq:purity4a}) and (\ref{eq:purity4b}), this term is
proportional to $[\var p(t)]^{-1/2}$. For $\alpha>1$ the
exponential decay therefore asymptotically crosses over to a
power-law decay with exponent $-1/2$.
 As discussed
before, for $\alpha<1$, ${\cal D}^2(t)$ itself already crosses
over from a stretched exponential to a power law with exponent
$-2\alpha$. As a consequence of the subdiffusive momentum
spreading, this intermediate power law is asymptotically replaced
by a slower power law with exponent $-\alpha/2$.

As a function of the inverse random time ${\cal N}(0,t)$, the
crossover to the asymptotic power-law of the saturation
background sets in at ${\cal N}(0,t)\approx t_c\ln t^*$. For
$\alpha>1$, this condition translates into a well-defined
crossover time $t\approx \overline{\tau}t_c\ln t^*$. For
$\alpha<1$, however, the nonergodicity of L\'evy noise entails
that the crossover is blurred.

\begin{figure}[t]
\epsfxsize=\columnwidth \epsffile{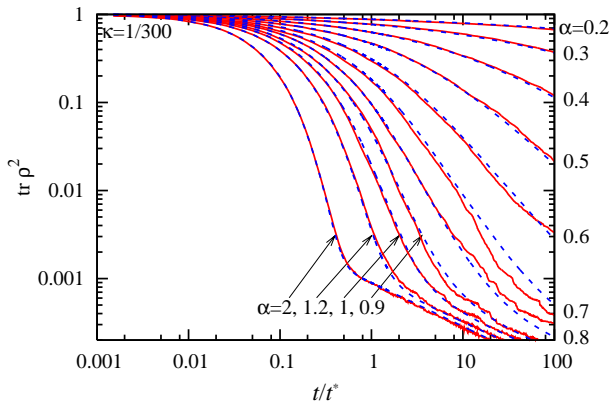} \caption{(color
online). Time dependence of the purity $\tr\rho^2$ of L{\'e}vy
kicked rotators for different noise exponents $\alpha$ and fixed
noise strength $\kappa=1/300$. The results of numerical
simulations (solid curves) are compared to the theoretical
predictions from Eq.\ (\ref{eq:purity}) (dashed
curves)\label{fig:7}}
\end{figure}

The numerical results presented in  Fig.~\ref{fig:7} confirm the
quantitative and qualitative predictions of Eq.~(\ref{eq:purity})
for the purity of the L{\'e}vy kicked rotator.  For $\alpha>1$ the
initial decay is exponential and the asymptotic decay is algebraic
with exponent $-1/2$. The background contribution is here clearly
visible and displays the predicted power-law decay.  For $\alpha <
1$ the decay is a stretched exponential which crosses over to an
algebraic decay with exponent $-2\alpha$. Because of this
significant slow-down of the decoherence, the eventual crossover
of the exponent to $-\alpha/2$ is only observed in the numerical
data for $\alpha \gtrsim 0.7$; for smaller values of $\alpha$, the
background contribution is  hardly discernible.

\begin{figure}
\epsfxsize=\columnwidth \epsffile{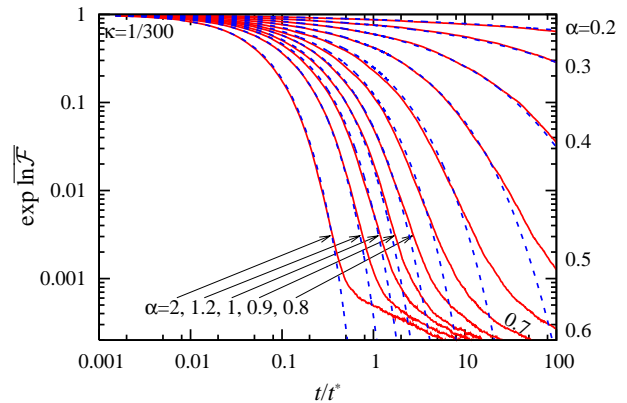}
\caption{\label{fig:8}(color online). Averaged logarithm of the
fidelity of L{\'e}vy kicked rotators for different noise exponents
$\alpha$ and fixed noise strength $\kappa=1/300$. The results of
numerical simulations (solid curves) are compared to the
theoretical predictions from Eq.\ (\ref{eq:ff1}) (dashed curves),
which does not account for the saturation background. }
\end{figure}

\subsection{Averaged logarithm of the fidelity}

For a system with complex quantum dynamics such as the kicked
rotator, it is often observed that the fidelity fluctuates
significantly from realization to realization \cite{silvestrov},
and from Fig.\ \ref{fig:7} it is clear that this tendency is
exacerbated for nonergodic noise. These fluctuations are
suppressed when one instead considers the averaged logarithm
$\overline{\ln{\cal F}(t)}$ of the fidelity.

By passing from the fidelity to its logarithm, the
multiplicative suppression of coherence per noise event is
converted into additive contributions. For initial and
intermediate times, the averaged logarithm hence  probes the
mean number of noise events in both noise realizations,
\begin{equation} \overline{{\cal N}(0,t)}+\overline{{\cal
N}'(0,t)}=2\int_0^t f(t')\,dt',
\end{equation}
where $f(t)$ is the sprinkling distribution. This leads to the
prediction
\begin{equation}
\overline{\ln{\cal F}(t)}=-\frac{2}{t_c}\int_0^t f(t')\,dt' \quad
\mbox{(short and intermediate times)}, \label{eq:ff1}
\end{equation}
while asymptotically a saturation background should set in (we do
not give an estimate of this background).

The validity of this prediction is confirmed by the numerical
results in Fig.\ \ref{fig:8}, which shows the averaged logarithm
in a representation that makes it readily comparable to the
results for the purity (Fig.\ \ref{fig:7}). There is good
agreement for small and intermediate time, until saturation sets
in, which is not incorporated in Eq.\  (\ref{eq:ff1}). Two
noteworthy observations in the comparison to the purity are first
the inequality $\exp(\overline{\ln {\cal F}})<\tr \rho^2$, and
secondly the fact that the statistical fluctuations are now
distinctively suppressed. This indicates that the logarithm of the
fidelity probes typical noise realizations, while the purity is
influenced by rare noise realizations in which coherence is
exceptionally well preserved.

\section{\label{sec:conclusions}Discussion and conclusions}
In this concluding section, we would like to situate our work in
the general context of the study of anomalous diffusion and
nonexponential relaxation. The investigation of diffusion in
complex environments enjoys a long history starting with the
seminal work of Scher and Montroll on dispersive transport in
amorphous semiconductors \cite{sch75}. Subsequently, extensive
work on non-Brownian motion and nonexponential decay induced by
L\'evy noise in space and/or in time in classical systems has been
carried out, most prominently using the powerful Continuous Time
Random Walk (CTRW) formalism (for a review see Ref.\
\cite{met00}). In the last few years, this line of research has
been extended to quantum systems. The novel aspect of these
developments is the possibility to study the interplay of complex
dynamics and quantum phenomena.  Thus the interaction with a
composite environment with extra degrees of freedom, and the
entanglement of the quantum system with the latter, has for
instance be shown to lead to power-law decay \cite{bud05}.  The
relaxation of a quantum two-level system subjected to stationary
and nonstationary power-law noise  has been respectively examined
in Refs.\ \cite{goy06} and \cite{sch05}. In the latter work,  the
presence of aging dephasing has been demonstrated. Additionally,
anomalous fast decoherence induced by spatial L\'evy noise,
stemming from a chaotic random-matrix environment, has been
investigated in Ref.\ \cite{lut02}.

Our approach in the present work has been to use the quantum
kicked rotator as a powerful tool to explore decoherence and its
interplay with complex quantum dynamics, in particular,
nonstationarity. The latter property is well-known from the
physics of disordered or glassy materials \cite{ric94}. Taking
advantage of the unique degree of tunability of the atom-optical
realization of the kicked rotator, we have put forward a way to
engineer a complex reservoir, providing full control over the
stationarity of the environment or the absence thereof.
Specifically, we have proposed to simulate the coupling to a
complex  reservoir with the help of L\'evy noise with variable
exponent. In this manner, we extend usual quantum reservoir
engineering in the spirit of the early studies of anomalous
diffusion in complex media based on the random walk concept. The
renewal process that we investigate can actually be regarded as
corresponding to a CTRW with power-law waiting time, but with a
renormalized diffusion range given by the localization length of
the kicked rotator.

At this point, it is appropriate to mention that anomalous
diffusion has been shown before to occur in quantum kicked
rotators subjected to {\em deterministic} aperiodic kicking. More
precisely, subballistic wave packet spreading induced by
quasiperiodic Fibonacci kick sequences has been established both
away and at resonance \cite{cas01,rom07}. The effect of L\'evy
noise at resonance has also been the subject of a recent study
\cite{rom07a}. An important difference to the present study is the
fact that these studies do not consider the simultaneous presence
of the periodic kicks in the intervals between the noise events.
As a consequence, the classical momentum dynamics for the
quasiperiodically kicked systems is subdiffusive, and hence does
not differ qualitatively from the quantum dynamics. The protocol
proposed in the present paper stipulates the presence of such
periodic kicks, which induce diffusive classical momentum
dynamics, while quantum mechanically they drive the system towards
localization. The ensuing qualitative difference between the
noiseless quantum and classical dynamics indeed provides the
terrain in which we explore the consequences of decoherence.

The theory of decoherence that we have developed in this paper
combines the properties of L\'evy noise, encoded in the generating
function of the number of noisy events (\ref{eq:mdef}), and a
random-phase approximation, which exploits the complex quantum
dynamics of the rotator appearing in the quasienergy transition
amplitudes (\ref{eq:transition}). It is worth emphasizing that it
is nonperturbative (with respect to time) and applies both to
stationary and nonstationary noise.

One of the central results of our paper, contained in Eq.\
(\ref{eq:MLp}), is the demonstration of a regime of nonexponential
loss of phase coherence when the noise is nonstationary (L{\'e}vy
exponent $\alpha<1$). This regime is characterized by an infinite
coherence time, indicating that classical behavior can never be
attained, even though the amplitude of the noise is dramatically
increased by several orders of magnitude. This remarkable property
finds its origin in the divergence of the mean waiting time
between random kicks and hence in the absence of a characteristic
time scale in the engineered reservoir.

 The existence of a regime
of slow decoherence has direct consequence for experimentally
accessible physical quantities of the kicked rotator. We have
first shown that the momentum spreading (\ref{eq:subdiff}) is
subdiffusive and that the momentum profile retains a double-sided
exponential shape, see Fig.\ \ref{fig:3}. These asymptotic
features are in stark contrast to those in the usual decoherence
regime, here realized for L{\'e}vy exponent $\alpha>1$, which is
characterized by normal diffusion and a Gaussian profile. The
momentum distribution is routinely measured in kicked rotator
experiments. In the atom-optical experiments \cite{amm98,kla98},
the variance of the momentum is accessible up to times of the
order of ten break times $t^*$, indicating that the subdiffusive
dynamics depicted in Fig.\ \ref{fig:5} is measurable with current
setups.

Two other important quantities that we have considered are the
fidelity and the purity (or average fidelity), which are more
traditional quantities for the characterization of decoherence.
Here again we have found clear signatures of the nonexponential
decoherence regime. In Eq.\ (\ref{eq:purity}) we have established
a direct connection between the purity on the one hand and the
decoherence factor and the inverse participation on the other. As
a consequence, we could show that for  $\alpha < 1$ the decay of
the purity evolves from a stretched exponential to an algebraic
decay, while $\alpha>1$ the  decay changes from exponential to
algebraic with exponent $-1/2$ (see Fig.\ \ref{fig:8}). The
fidelity can be directly measured in echo experiments, see for
example Ref.\ \cite{sch05}, however to our knowledge such a
measurement has not yet been performed in kicked rotator
experiments. We also mention that we have found a direct
relationship (\ref{eq:ff1}) between the averaged logarithm of the
fidelity  and the sprinkling distribution of the renewal process.

Our concluding message is  that reservoir engineering can be
extended to mimic the coupling to  complex nonstationary
environments, allowing to control in a precise manner the loss of
phase coherence and reveal hitherto unexplored decoherence
scenarios.

This work was supported by the European Commission, Marie Curie
Excellence Grant MEXT-CT-2005-023778 (Nanoelectrophotonics),
the Emmy Noether program of the DFG, contract LU1382/1-1, and the
cluster of excellence Nanosystems Initiative Munich (NIM).

\end{document}